# The effects of change decomposition on code review—a controlled experiment

Marco di Biase[1,2], Magiel Bruntink[2], Arie van Deursen[1] and Alberto Bacchelli[3]

[1] Delft University of Technology, Delft, The Netherlands
[2] Software Improvement Group, Amsterdam, The Netherlands
[3] University of Zurich, Zurich, Switzerland

## ABSTRACT

**Background:** Code review is a cognitively demanding and time-consuming process. Previous qualitative studies hinted at how decomposing change sets into multiple yet internally coherent ones would improve the reviewing process. So far, literature provided no quantitative analysis of this hypothesis.
**Aims:** (1) Quantitatively measure the effects of change decomposition on the outcome of code review (in terms of number of found defects, wrongly reported issues, suggested improvements, time, and understanding); (2) Qualitatively analyze how subjects approach the review and navigate the code, building knowledge and addressing existing issues, in large vs. decomposed changes.
**Method:** Controlled experiment using the pull-based development model involving 28 software developers among professionals and graduate students.
**Results:** Change decomposition leads to fewer wrongly reported issues, influences how subjects approach and conduct the review activity (by increasing context-seeking), yet impacts neither understanding the change rationale nor the number of found defects.
**Conclusions:** Change decomposition reduces the noise for subsequent data analyses but also significantly supports the tasks of the developers in charge of reviewing the changes. As such, commits belonging to different concepts should be separated, adopting this as a best practice in software engineering.

# INTRODUCTION

Code review is one among the activities performed by software teams to check code quality, with the purpose of identifying issues and shortcomings (*Bacchelli & Bird, 2013*). Nowadays, reviews are mostly performed in an iterative, informal, change- and tool-based fashion, also known as modern code review (MCR) (*Cohen, 2010*). Software development teams, both in open-source and industry, employ MCR to check code changes before being integrated in to their codebases (*Rigby & Bird, 2013*). Past research has provided evidence that MCR is associated with improved key software quality aspects, such as maintainability (*Morales, McIntosh & Khomh, 2015*) and security (*di Biase, Bruntink & Bacchelli, 2016*), as well as with less defects (*McIntosh et al., 2016*).

Reviewing a source code change is a cognitively demanding process. Researchers provided evidence that understanding the code change under review is among the most challenging tasks for reviewers (*Bacchelli & Bird, 2013*). In this light, past studies have argued that code changes that—at the same time—address multiple, possibly unrelated concerns (also known as *noisy* (*Murphy-Hill, Parnin & Black, 2012*) or *tangled changes* (*Herzig & Zeller, 2013*)) can hinder the review process (*Herzig & Zeller, 2013*; *Kirinuki et al., 2014*), by increasing the cognitive load for reviewers. Indeed, it is reasonable to think that grasping the rationale behind a change that spans multiple concepts in a system requires more effort than the same patch committed separately. Moreover, the noise could put a reviewer on a wrong track, thus leading to missing defects (*false negatives*) or to raising unfounded issues in sound code (*false positives* in this paper).

Qualitative studies reported that professional developers perceive tangled code changes as problematic and asked for tools to automatically decompose them (*Tao et al., 2012*; *Barnett et al., 2015*). Accordingly, change untangling mechanisms have been proposed (*Tao & Kim, 2015*; *Dias et al., 2015*; *Barnett et al., 2015*).

Although such tools are expectedly useful, the effects of change decomposition on review is an open research problem. *Tao & Kim (2015)* presented the earliest and most relevant results in this area, showing that change decomposition allows practitioners to achieve their tasks better in a similar amount of time.

In this paper, we continue on this research line and focus on evaluating the effects of change decomposition on code review. We aim at answering questions, such as: Is change decomposition beneficial for understanding the rationale of the change? Does it have an impact on the number/types of issues raised? Are there differences in time to review? Are there variations with respect to defect lifetime?

To this end, we designed a controlled experiment focusing on pull requests, a widespread approach to submit and review changes (*Gousios et al., 2015*). Our work investigates whether the results from *Tao & Kim (2015)* can be replicated, and extend the knowledge on the topic. With a Java system as a subject, we asked 28 software developers among professionals and graduate students to review a refactoring and a new feature (according to professional developers (*Tao et al., 2012*), these are the most difficult to review when tangled). We measure how the partitioning vs. non-partitioning of the changes impacts defects found, false positive issues, suggested improvements, time to review, and understanding the change rationale. We also perform qualitative observations on how subjects conduct the review and address defects or raise false positives, in the two scenarios.

This paper makes the following contributions:

- The design of an asynchronous controlled experiment to assess the benefits of change decomposition in code review using pull requests, available for replication (*di Biase et al., 2018*);
- Empirical evidence that change decomposition in the pull-based review environment leads to fewer false positives.

# RELATED WORK

Several studies explored tangled changes and concern separation in code reviews. *Tao et al. (2012)* investigated the role of understanding code changes during the software development process, exploring practitioners' needs. Their study outlined that grasping the rationale when dealing with the process of code review is indispensable. Moreover, to understand a composite change, it is useful to break it into smaller ones each concerning a single issue. *Rigby et al. (2014)* empirically studied the peer review process for six large, mature OSS projects, showing that small change size is essential to the more fine-grained style of peer review. *Kirinuki et al. (2014)* provided evidence about problems with the presence of multiple concepts in a single code change. They showed that these are unsuitable for merging code from different branches, and that tangled changes are different to review because practitioners have to seek the changes for the specified task in the commit.

Regarding empirical controlled experiments on the topic of code reviews, the most relevant work is by *Uwano et al. (2006)*. They used an eye-tracker to characterize the performance of subjects reviewing source code. Their experimentation environment enabled them to identify a pattern called *scan*, consisting of the reviewer reading the entire code before investigating the details of each line. In addition, their qualitative analysis found that participants who did not spend more time during the *scan* took more time to find defects. Uwano's experiment was replicated by *Sharif, Falcone & Maletic (2012)*. Their results indicated that the longer participants spent in the *scan*, the quicker they were able to find the defect. Conversely, review performance decreases when participants did not spend sufficient time on the scan, because they find irrelevant lines. Recently, *Baum, Schneider & Bacchelli (2019)* highlighted how performance in code review is significantly higher when code changes are small, whereas complex and longer changes lead to lower review effectiveness.

Even if MCR is now a mainstream process, adopted in both open source and industrial projects, we found only two studies on change partitioning and its benefits for code review. The work by *Barnett et al. (2015)* analyzed the usefulness of an automatic technique for decomposing changesets. They found a positive association between change decomposition and the level of understanding of the changesets. According to their results, this would help time to review as the different contexts are separated. *Tao & Kim (2015)* proposed a heuristic-based approach to decompose changeset with multiple concepts. They conducted a user study with students investigating whether their untangling approach affected the time and the correctness in performing review-related tasks. Results were promising: Participants completed the tasks better with untangled changes in a similar amount of time. In spite of the innovative techniques they proposed to untangle code changes and in these promising results, the evaluation of effects of change decomposition was preliminary.

In contrast, our research focuses on setting up and running an experiment to empirically assess the benefits of change decomposition for the process of code review, rather than evaluating the performances of an approach.

## MOTIVATION AND RESEARCH OBJECTIVES

In this section, we present the context of our work and the research questions.

### Experiment definition and context

Our analysis of the literature showed that there is only preliminary empirical evidence on how code review decomposition affects its outcomes, its change understanding, time to completion, effectiveness (i.e., number of defects found), false positives (issues mistakenly identified as defect by the reviewer), and suggested improvements. This lack of empirical evidence motivates us in setting up a controlled experiment, exploiting the popular pull-based development model, to assess the conjecture that a proper separation of concerns in code review is beneficial to the efficiency and effectiveness of the review.

Pull requests feature asynchronous, tool-based activities in the bigger scope of pull-based software development (*Gousios, Pinzger & Van Deursen, 2014*). The pull-based software process features a distributed environment where changes to a system are proposed through patch submissions that are pulled and merged locally, rather than being directly pushed to a central repository.

Pull requests are the way contributors submit changes for review in GitHub. Change acceptance has to be granted by other team members called integrators (*Gousios et al., 2015*). They have the crucial role of managing and integrating contributions and are responsible for inspecting the changes for functional and non-functional requirements. A total of 80% of integrators use pull requests as the means to review changes proposed to a system (*Gousios et al., 2015*).

In the context of distributed software development and change integration, GitHub is one of the most popular code hosting sites with support for pull-based development. GitHub pull requests contain a branch from which changes are compared by an automatic discovery of commits to be merged. Changes are then reviewed online. If further changes are requested, the pull request can be updated with new commits to address the comments. The inspection can be repeated and, when the patch set fits the requirements, the pull request can be merged to the master branch.

### Research questions

The motivation behind MCR is to find defects and improve code quality (*Bacchelli & Bird, 2013*). We are interested in checking if reviewers are able to address *defects* (referred in this paper as *effectiveness*). Furthermore, we focus on comments pointing out *false positives* (wrongly reported defects), and *suggested improvements* (non-critical non-functional issues such as suggested refactorings). Suggested improvements highlight reviewer participation (*McIntosh et al., 2014*) and these comments are generally considered very useful (*Bosu, Greiler & Bird, 2015*). Our first research question is:

> **RQ1.** Do tangled pull requests influence *effectiveness* (i.e., number of defects found), *false positives*, and *suggested improvements* of reviewers, when compared to untangled pull requests?

Based on the first research question, we formulate the following null-hypotheses for (statistical) testing:

Tangled pull requests do not reduce:

$H_{0e}$ the effectiveness of the reviewers during peer-review

$H_{0f}$ the false positives detected by the reviewers during peer-review

$H_{0c}$ the suggested improvements written by the reviewers during peer-review

Given the structure and the settings of our experimentation, we can also measure the time spent on review activity and defect lifetime. Thus, our next research question is:

**RQ2.** Do tangled pull requests influence the time necessary for a review and defect lifetime, when compared to untangled pull requests?

For the second research question, we formulate the following null-hypotheses:

Tangled pull requests do not reduce:

$H_{0t1}$ time to review

$H_{0t2}$ defect lifetime

Further details on how we measure time and define defect lifetime are described in the section "Outcome Measurements".

In our study, we aim to measure whether change decomposition has an effect on understanding the rationale of the change under review. Understanding the rationale is the most important information need when analyzing a change, according to professional software developers (*Tao et al., 2012*). As such, the question we set to answer is:

**RQ3.** Do tangled pull requests influence the reviewers' understanding of the change rationale, when compared to untangled ones?

For our third research question, we test the following null-hypotheses:

Tangled pull requests do not reduce:

$H_{0u}$ change-understanding of reviewers during peer-review when compared to untangled pull requests

Finally, we qualitatively investigate how participants individually perform the review to understand how they address defects or potentially raise false positives. Our last research question is then:

**RQ4.** What are the differences in patterns and features used between reviews of tangled and untangled pull requests?

**Table 1** Descriptive data of the subject participants.

| Group | # of subjects | | Role | FTE experience | | Reviews per week | | | |
|---|---|---|---|---|---|---|---|---|---|
| | Total | with system knowledge | | μ | | per role | | per group | |
| | | | | | | μ | | μ | |
| Control (tangled changes) | 6 | 2 (33%) | SW developer | 4.3 | 4.8 | 4.8 | 3.3 | 3.6 | 3.6 |
| | 3 | 1 (33%) | PhD student | 5.0 | 2.9 | 3.0 | 2.9 | | |
| | 5 | 3 (60%) | MSc student | 2.2 | 0.7 | 2.6 | 3.8 | | |
| Treatment (untangled changes) | 6 | 2 (33%) | SW developer | 4.8 | 2.9 | 3.3 | 3.4 | 4.0 | 6.4 |
| | 3 | 1 (33%) | PhD student | 6.0 | 6.6 | 2.0 | 0.8 | | |
| | 5 | 3 (60%) | MSc student | 2.2 | 1.1 | 6.0 | 9.0 | | |

# EXPERIMENTAL DESIGN AND METHOD

In this section, we detail how we designed the experiment and the research method that we followed.

## Object system chosen for the experiment

The system that was used for reviews in the experiment is JPacman, an open-source Java system available on GitHub (https://github.com/SERG-Delft/jpacman-framework) that emulates a popular arcade game used at Delft University of Technology to teach software testing.

The system has about 3,000 lines of code and was selected because a more complex and larger project would require participants to grasp the rationale of a more elaborate system. In addition, the training phase required for the experiment would imply hours of effort, increasing the consequent fatigue that participants might experience. In the end, the experiment targets assessing differences in review partitioning and is tailored for a process rather than a product.

## Recruiting of the subject participants

The study was conducted with 28 participants recruited by means of convenience sampling (Wohlin et al., 2012) among experienced and professional software developers, PhD, and MSc students.[1] They were drawn from a population sample that volunteered to participate. The voluntary nature of participation implies the consent to use data gathered in the context of this study. Software developers belong to three software companies, PhD students belong to three universities, and MSc students to different faculties despite being from the Delft University of Technology. We involved as many different roles as possible to have a larger sample for our study and increase its external validity. Using a questionnaire, we asked development experience, language-specific skills, and review experience as number of reviews per week. We also included a question that asked if a participant knew the source code of the game. Table 1 reports the results of the questionnaire, which are used to characterize our population and to identify key attributes of each subject participant.

[1] Delft University of Technology Human Research Committee approved our study with IRB approval #578. University of Zurich authorized the research with IRB approval #2018-024.

## Monitoring vs. realism

In line with the nature of pull-based software development and its peer review with pull requests, we designed the experimentation phase to be executed asynchronously. This implies that participants could run the experiment when and where they felt most comfortable, with no explicit constraints for place, time or equipment.

With this choice, we purposefully gave up some degree of control to increase realism. Having a more strictly controlled experimental environment would not replicate the usual way of running such tasks (i.e., asynchronous and informal). Besides, an experiment run synchronously in a laboratory would still raise some control challenges: It might be distracting for some participants, or even induce some *follow the crowd* behavior, thus leading to people rushing to finish their tasks.

To regain some degree of control, participants ran all the tasks in a provided virtual machine available in our replication package (*di Biase et al., 2018*). Moreover, we recorded the screencast of the experiment, therefore not leaving space to misaligned results and mitigating issues of incorrect interpretation. Subjects were provided with instructions on how to use the virtual machine, but no time window was set.

## Independent variable, group assignment, and duration

The independent variable of our study is change decomposition in pull requests. We split our subjects between a *control* group and a *treatment* group: The control group received one pull request containing a single commit with all the changes tied together; the treatment group received two pull requests with changes separated according to a logical decomposition. Our choice of using only two pull requests instead of a larger number is mainly dictated by the limited time participants were allotted for the experiment, and the possibly increased influence of distractions. Changes spanning a greater part of the codebase require additional expertise, knowledge, and focus, which reviewers might lack. Extensive literature in psychology (*Shiffrin, 1988*; *Wickens, 1991*; *Cowan, 1998*; *James, 2013*) reports that cognitive resources such as attention are finite. Since complex tasks like reviewing code drain such resources, the effectiveness of the measured outcomes will be negatively impacted by a longer duration.

Participants were randomly assigned to either the control group or the treatment using strata based on experience as developers and previous knowledge. Previous research has shown that these factors have an impact on review outcome (*Rigby et al., 2012*; *Bosu, Greiler & Bird, 2015*). Developers who previously made changes to files to be reviewed had a higher proportion of useful comments.

All subjects were asked to run the experiment in a single session so that external distracting factors could be eliminated as much as possible. If a participant needed a pause, the pause is considered and excluded from the final result as we measure and monitor for periods of inactivity. We seek to reduce the impact of fatigue by limiting the expected time required for the experiment to an average of 60 min; this value is closer to the minimum rather than the median for similar experiments (*Ko, LaToza & Burnett, 2015*). As stated before, though, we did not suggest or force any strict limit on the duration of the experiment to the ends of replicating the code review informal scenario. No learning effect is present as every participant runned the experiment only once.

## Pilot experiments

We ran two pilot experiments to assess the settings. The first subject (a developer with five FTE[2] years of experience) took too long to complete the training and showed some issues with the virtual machine. Consequently, we restructured the training phase addressing the potential environment issues in the material provided to participants. The second subject (a MSc student with little experience) successfully completed the experiment in 50 min with no issues. Both pilot experiments were executed asynchronously in the same way as the actual experiment.

## Tasks of the experiment

The participants were asked to conduct the following four tasks. Further details are available in the online appendix (*di Biase et al., 2018*).

### Preparing the environment

Participants were given precise and detailed instructions on how to set-up the environment for running the experiment. These entailed installing the virtual machine, setting up the recording of the screen during the experiment, and troubleshooting common problems, such as network or screen resolution issues.

### Training the participants

Before starting with the review phase, we first ensured that the participants were sufficiently familiar with the system. It is likely that the participants had never seen the codebase before: this situation would limit the realism of the subsequent review task.

To train our participants we asked subjects to implement three different features in the system:

1. Change the way the player moves on the board game, using different keys,
2. check if the game board has null squares (a board is made of multiple squares) and perform this check when the board is created, and
3. implement a new enemy in the game, with similar artificial intelligence to another enemy but different parameters.

This learning by doing approach is expected to have higher effectiveness than providing training material to participants (*Slavin, 1987*). By definition, this approach is a method of instruction where the focus is on the role of feedback in learning. The desired features required change across the system's codebase. The third feature to be implemented targeted the classes and components of the game that would be object of the review tasks. The choice of using this feature as the last one is to progressively increment the level of difficulty.

No time window was given to participants, aiming for a more realistic scenario. As explicitly mentioned in the provided instructions, participants were allowed to use any source for retrieving information about something they did not know. This was permitted as the study does not want to assess skills in implementing some functionality in a programming language. The only limitation is that the participants must use the tools within the virtual machine.

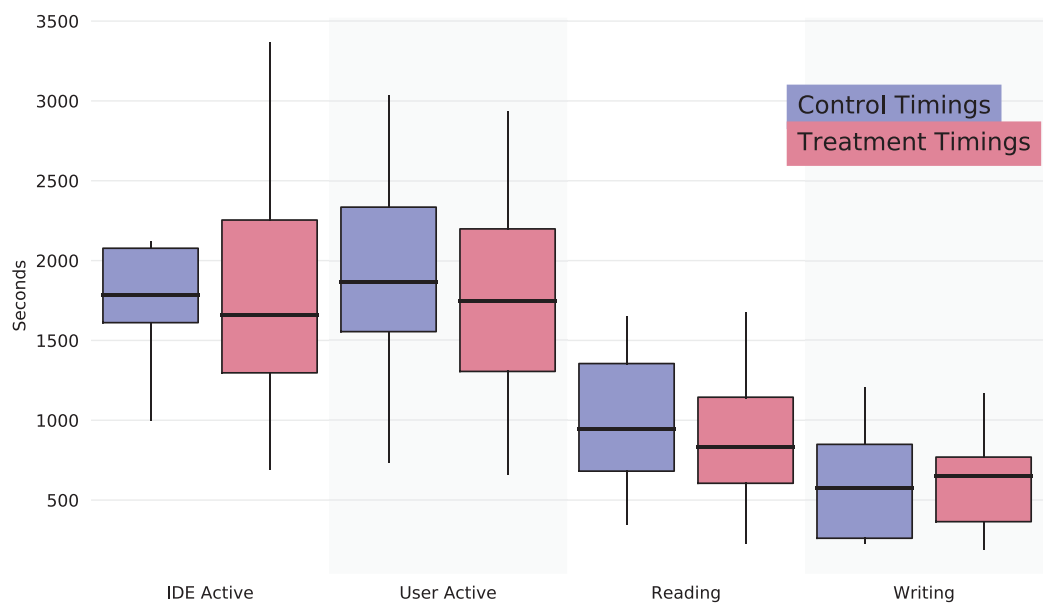

The virtual machine provided the participants with the Eclipse Java IDE. The setup already had the project imported in Eclipse's workspace. We used a plugin in Eclipse, WatchDog (*Beller et al., 2015*), to monitor development activity. With this plugin, we measured how much time participants spent reading, typing, or using the IDE. The purpose was to quantify the time to understand code among participants and whether this relates to a different outcome in the following phases. Results for this phase are shown in Fig. 1, which contains boxplots depicting the data. It shows that there is no significant difference between the two groups. We retrieve the non-statistical significance by performing Mann–Whitney $U$-tests on the four variables in Fig. 1, with the following $p$-values: IDE active: $p$-value = 0.98, User Active: $p$-value = 0.80, Reading: $p$-value = 0.73, Writing: $p$-value = 0.73.

### Perform code review on proposed change(s)

Participants were asked to review two changes made to the system:

1. the implementation of the artificial intelligence for one of the enemies
2. the refactoring of a method in all enemy classes (moving its logic to the parent class).

These changes can be inspected in the online appendix (*di Biase et al., 2018*) and have been chosen to meet the same criteria used by *Herzig, Just & Zeller (2016)* when choosing tangled changes. Changes proposed can be classified as *refactoring* and *enhancement*. Previous literature gave insight as to how these two kinds of changes, when tangled together, are the hardest to review (*Tao et al., 2012*). Although recent research proposed a theory for the optimal ordering of code changes in a review (*Baum, Schneider &*

*Bacchelli, 2017*), we used the default ordering and presentation provided by GitHub, because it is the de-facto standard. Changesets were included in pull requests on private GitHub repositories so that participants performed the tasks in a real-world review environment. Pull requests had identical descriptions for both the control and the treatment, with no additional information except their descriptive title. While research showed that a short description may lead to poor review participation (*Thongtanunam et al., 2017*), this does not apply to our experiment as there is no interaction among subjects.

Subjects were instructed to understand the change and check its functional correctness. We asked the participants to comment on the pull request(s) if they found any problem in the code, such as any functional error related to correctness and issues with code quality. The changes proposed had three different functional issues that were intentionally injected into the source code. Participants could see the source code of the whole project in case they needed more context, but only through GitHub's browser-based UI.

The size of the changeset was around 100 lines of code and it involved seven files. *Gousios, Pinzger & Van Deursen (2014)* showed that the number of total lines changed by pull requests is on average less than 500, with a median of 20. Thus, the number of lines of the changeset used in this study is between the median and the average. Our changeset size is also consistent with recent research which found that code review is highly effective on smaller changes (*Baum, Schneider & Bacchelli, 2019*) and rewiewability can be empirically defined through several factors, one being change size (*Ram et al., 2018*)

### Post-experiment questionnaire

In the last phase participants were asked to answer a post-experiment questionnaire. Questions are showed in the section "Results," RQ3: Q1–Q4 were about change-understanding, while Q5–Q12 involved subjects' opinions about changeset comprehension and its correctness, rationale, understanding, etc. Q5–Q12 were a summary of interesting aspects that developers need to grasp in a code change, as mentioned in the study of *Tao et al. (2012)*. The answers must be provided in a Likert scale (*Oppenheim, 2000*) ranging from "Strongly disagree" (1) to "Strongly agree" (5).

## Outcome measurements
### Effectiveness, false positives, suggested improvements

Subjects were asked to comment a pull request in the pull request discussion or in-line comment in a commit belonging to that pull request. The number of comments addressing functional issues was counted as the effectiveness. At the same time, we also measured false positives (i.e., comments in pull request that do not address a real issue in the code) and suggested improvements (i.e., remarks on other non-critical non-functional issues).

We distinguished suggested improvements and false positives from the comments that correctly addressed an issue because the three functional defects were intentionally put in the source code. Comments that did not directly and correctly tackle one of these three issues were classified either as false positives or suggested improvements. They were classified by the first author by looking at the description provided by the subject. A correctly identified issue needs to highlight the problem, and optionally provide a short description.

### Time

Having the screencast of the whole experiment, as well as using tools that give time measures, we gathered the following measurements:

- Time for Task 2, in particular:

  - total time Eclipse is [opened/active]
  - total time the user is [active/reading/typing];
    as collected by WatchDog (section "Tasks of the Experiment").

- Total net time for Task 3, defined as from when the subject opens a pull request until when (s)he completes the review, purged of eventual breaks.
- Defect lifetime, defined as the period during which a defect continues to exist. It is measured from the moment the subject opens a pull request to when (s)he writes a comment that correctly identifies the issue. For the case of multiple comments on the same pull request, this is the time between finishing with one defect and addressing the next. A similar measure was previously used by *Prechelt & Tichy (1998)*.

  All the above measures are collected in seconds elapsed.

### Change-understanding

In this experiment, change understanding was measured by means of a questionnaire submitted to participants post review activity, as mentioned in Task 4 in the section "Tasks of the Experiment." Questions are shown in the section "Results," RQ3, from Q1 to Q4. Its aim is to evaluate differences in change-understanding. A similar technique was used by *Binkley et al. (2013)*.

### Final survey

Lastly, participants were asked to give their opinion on statements targeting the perception of correctness, understanding, rationale, logical partitioning of the changeset, difficulty in navigating the changeset in the pull request, comprehensibility, and the structure of the changes. This phase, as well as the previous one, was included in Task 4, corresponding to questions Q5–Q12 (section "Results," RQ3). Results were given on a Likert scale from "Strongly disagree" (1) to "Strongly agree" (5) (*Oppenheim, 2000*), reported as mean, median and standard deviation over the two groups, and tested for statistical significance with the Mann–Whitney $U$-test.

## Research method for RQ4

For our last research question, we aimed to build some initial hypothesis to explain the results from the previous research questions. We sought what actions and patterns led a reviewer in finding an issue or raising false positive, as well as other comments. This method was applied only to the review phase, without analyzing actions and patterns concerning the training phase. The method to map actions to concepts started by annotating the screencasts retrieved after the conclusion of the experimental phase. Subjects performed a series of actions that defined and characterized both the outcome and the execution of the review. The first author inserted notes regarding actions performed by participants to build a knowledge base of steps (i.e., participant opens `fileName`, participant uses GitHub search box with the `keyword`, etc.).

Using the methodology for qualitative content analysis delineated by *Schreier (2013)*, we firstly defined the coding frame. Our goal was to characterize the review activity based on patterns and behaviors. As previous studies already tackled this problem and came up with reliable categories, we used the investigations by *Tao et al. (2012)* and *Sillito, Murphy & De Volder (2006)* as the base for our frame. We used the concepts from *Tao et al. (2012)* regarding *Information needs for reasoning and assessing the change* and *Exploring the context and impact of the change*, as well as the *Initial focus points* and *Building on initial focus points* steps from *Sillito, Murphy & De Volder (2006)*.

To code the transcriptions, we used the deductive category application, resembling the data-driven content analysis technique by *Mayring (2000)*. We read the material transcribed, checking whether a concept covers that action transcribed (e.g., participant opens file `fileName` so that (s)he is looking for context). We grouped actions covered by the same concept (e.g., a participant opens three files, but always for context purpose) and continued until we built a pattern that led to a specific outcome (i.e., addressing a defect or a false positive). We split the patterns according to their concept ordering such that those that led to more defects found or false positive issues were visible.

# THREATS TO VALIDITY AND LIMITATIONS

## Internal validity

The sample size comprised in our experiment poses an inherent threat to the internal validity of our experiment. Furthermore, using a different experimental strategy (e.g., that used by *Tao & Kim (2015)*) would remove personal performance biases, while causing a measurable learning effect. In fact, *Wohlin et al. (2012)* state that "*due to learning effects, a human subject cannot apply two methods to the same piece of code.*" This would result in affecting the study goals and construct validity. In addition, the design and asynchronous execution of the experimental phase creates uncertainty regarding possible external interactions. We could not control random changes in the experimental setting, and this translates to possible disturbances coming from the surrounding environment, that could cause skewed results.

Moreover, our experiment settings could not control if participants interacted among them, despite participants did not have any information about each other.

Regarding the statistical regression (*Wohlin et al., 2012*), tests used in our study not performed with the Bonferroni correction, following the advice by Perneger: "*Adjustments are, at best, unnecessary and, at worst, deleterious to sound statistical inference*" (*Perneger, 1998*). Other corrections such as the false discovery rate (FDR) are also not suited for our study. The de-facto standard to perform the FDR correction is the Benjamini–Hochberg (BH) (*Benjamini & Hochberg, 1995*). The BH, though useful when dealing with large numbers of $p$-values (e.g., 100), needs careful adjustment of a threshold to detect false positives. The number of statistical tests performed in our study is small enough to warrant not applying FDR or other significance corrections.

## Construct validity

Relatively to the restricted generalizability across constructs (*Wohlin et al., 2012*), in our experiment, we uniquely aim to measure the values presented in the section "Outcome Measurements." The treatment might influence direct values we measure, but it could potentially cause negative effects on concepts that our study does not capture. Additionally, we acknowledge threats regarding the time measures taken by the first author regarding RQ2. Clearly, manual measures are suboptimal, that were adopted to avoid participants having to perform such measures themselves.

When running an experiment, participants might try to guess what is the purpose of the experimentation phase. Therefore, we could not control their behavior based on the guesses that either positively or negatively affected the outcome.

Threats to construct validity are connected to designing the questionnaires used for RQ3, despite designed using standard ways and scales (*Oppenheim, 2000*). Finally, threats connected to the manual annotation of the screencasts recorded and analyzed by the first author could lead to misinterpreted or misclassified actions performed by the participants in our experiment.

## External validity

Threats to external validity for this experiment concern the selection of participants to the experimentation phase. Volunteers selected with convenience sampling could have an impact on the generalizability of results, which we tried to mitigate by sampling multiple roles for the task. If the group is very heterogeneous, there is a risk that the variation due to individual differences is larger than due to the treatment (*Cook & Campbell, 1979*).

Furthermore, we acknowledge and discuss the possible threat regarding the system selection for the experimental phase. Naturally, the system used is not fully representative of a real-world scenario. Our choice, however, as explained in the section "Object System Chosen for the Experiment," aims to reduce the training phase effort required from participants and to encourage the completion of the experiment. Despite research empirically showed that small code changes are easier to review (*Ram et al., 2018*) and current industrial practice reports reviews being almost always done on very small changesets (*Sadowski et al., 2018*), the external validity of our study is influenced by the size of the changes under review and number of pull requests in the experimental setup. Lack of empirical studies that provide an initial reference on the size of tangled changesets left us unable to address such threats. Future research should provide empirical evidence about the average tangled change, as well as the impact of larger changeset or number of pull requests on the results of our experiment.

Finally, our experiment was designed considering only a single programming language, using the pull-based methodology to review and accept the changes proposed using GitHub as a platform. Therefore, threats for our experiment are related to mono-operation and mono-method bias (*Wohlin et al., 2012*).

**Table 2** RQ1—number of defects found (effectiveness), false positives and suggested improvements.

| | Group | # of subjects | Total | Median | Mean | | Confidence interval 95% | *p*-value |
|---|---|---|---|---|---|---|---|---|
| Effectiveness (defects found) | Control | 14 | 20 | 1.0 | 1.42 | 0.72 | [0–2.85] | 0.6 |
| | Treatment | 14 | 17 | 1.0 | 1.21 | 0.77 | [0–2.72] | |
| False positives | Control | 14 | 6 | 0 | 0.42 | 0.5 | [0–1.40] | **0.03** |
| | Treatment | 14 | 1 | 0 | 0.07 | 0.25 | [0–0.57] | |
| Suggested improvements | Control | 14 | 7 | 0 | 0.5 | 0.62 | [0–1.22] | 0.4 |
| | Treatment | 14 | 19 | 1.0 | 1.36 | 1.84 | [0–5.03] | |

**Note:**
Statistically significant *p*-values in boldface.

# RESULTS

This section presents the results to the four research questions we introduced in the section "Research Questions".

## RQ1. Effectiveness, false positives, and suggestions

For our first research question, descriptive statistics about results are shown in Table 2. It contains data about effectiveness of participants (i.e., correct number of issues addressed), false positives, and number of suggested improvements. Given the sample size, we applied a non-parametric test and performed a Mann–Whitney *U*-test to test for differences between the control and the treatment group. This test, unlike a *t*-test, does not require the assumption of a normal distribution of the samples. Results of the statistical test are intended to be significant for a confidence level of 95%.

Results indicate a significant difference between the control and the treatment group regarding the number of false positives, with a *p*-value of 0.03. On the contrary, there is no statistically significant difference regarding the number of defects found (effectiveness) and number of suggested improvements.

The example of a false positive is when one of the subjects of the control group writes: "*This doesn't sound correct to me. Might want to fix the for, as the variable* `varName` *is never used.*" This is not a defect, as `varName` is used to check how many times the for-statement has to be executed, despite not being used inside the statement. This is also written in a code comment. Another false positive example is provided from a participant in the control group who, reading the *refactoring* proposed by the changeset under review, writes: "*The method* `methodName` *is used only in Class* `ClassName`, *so fix this.*" This is not a defect as the same `methodName` is used by the other classes in the hierarchy. As such, we can reject only the null hypothesis $H_{0f}$ regarding the false positives, while we cannot provide statistically significant evidence about the other two variables tested in $H_{0e}$ and $H_{0c}$.

The statistical significance alone for the false positives does not provide a measure to the actual impact of the treatment. To measure the effect size of the factor over the dependent variable we chose the Cliff's Delta (*Cliff, 1993*), a non-parametric measure for effect size. The calculation is given by comparing each of the scores in one group to each of the scores in the other, with the following formula: $\delta = \frac{\#(x_1 > x_2) - \#(x_1 < x_2)}{n_1 n_2}$ where $x_1$, $x_2$ are values for the two groups and $n_1$, $n_2$ are their sample size. For data with skewed

**Table 3** RQ2—review time, first and second defect lifetime.

| | Group | # of subjects | Median | Mean | | Confidence Interval 95% | *p*-value |
|---|---|---|---|---|---|---|---|
| Review net time | Control (tangled changes) | 14 | 831 | 853 | 385 | [99–1,607] | 0.66 |
| | Treatment (untangled changes) | 14 | 759 | 802 | 337 | [140–1,463] | |
| 1st defect lifetime | Control | 11 | 304 | 349 | 174 | [8–691] | 0.79 |
| | Treatment | 11 | 297 | 301 | 109 | [86–516] | |
| 2nd defect lifetime | Control | 6 | 222 | 263 | 149 | [0–555] | 0.17 |
| | Treatment | 6 | 375 | 388 | 122 | [148–657] | |

Note:
Measurements in seconds elapsed.

marginal distribution it is a more robust measure if compared to Cohen standardized effect size (*Cohen, 1992*). The computed value shows a positive (i.e., tangled pull requests lead to more false positives) effect size ($\delta = 0.36$), revealing a medium effect. The effect size is considered negligible for $|\delta| < 0.147$, small for $|\delta| < 0.33$, medium for $|\delta| < 0.474$, large otherwise (*Romano et al., 2006*).

> **Result 1:** *Untangled pull requests (treatment) lead to fewer false positives with a statistically significant, medium size effect.*

Given the presence of *suggested improvements* in our results, we found that the control group writes in total seven, while the participants in the treatment write 19. This difference is interesting, calling for further classification of the suggestions. For the control group, participants wrote respectively three improvements regarding *code readability*, two concerning *functional checks*, one regarding *understanding* of source code and one regarding other code issues. For the treatment group, we classified five suggestions for *code readability*, eight for *functional checks* and seven for *maintainability*. Although subjects have been explicitly given the goal to find and comment exclusively functional issues (section "Tasks of the Experiment"), they wrote these suggestions spontaneously. The suggested improvements are included in the online appendix (*di Biase et al., 2018*) along with their classification.

## RQ2. Review time and defect lifetime

To answer RQ2, we measured and analyzed the time subjects took to review the pull requests, as well as the amount of time they used to fix each of the issues present. Descriptive statistics about results for our second research question are shown in Table 3. It contains data about the time participants used to review the patch, completed by the measurements of how much they took to fix respectively two of the three issues present in the changeset. All measures are in seconds. We exclude data relatively to the third defect as only one participant detected it. To perform the data analysis, we used the same statistical means described for the previous research question. When computing the review net time used by the subjects, results show an insignificant difference, thus we are not able to reject null-hypothesis $H_{0t1}$. This indicates that the average case of the treatment group takes the same time to deliver the review, despite having two pull

**Table 4** RQ3—Post-experiment questionnaire.

Questions on understanding the rationale of the changeset

The purpose of this changeset entails ...

| | |
|---|---|
| Q1 | ... changing a method for the enemy AI |
| Q2 | ... the refactoring of some methods |
| Q3 | ... changing the game UI panel |
| Q4 | ... changing some method signature |

Questions on participant's perception on the changeset

| | |
|---|---|
| Q5 | The changeset was functionally correct |
| Q6 | I found no difficulty in understanding the changeset |
| Q7 | The rationale of this changeset was perfectly clear |
| Q8* | The changeset a logical separation of concerns |
| Q9 | Navigating the changeset was hard |
| Q10* | The relations among the changes were well structured |
| Q11 | The changeset was comprehensible |
| Q12* | Code changes were spanning too many features |

**Note:**
Questions with * have $p < 0.05$.

requests to deal with instead of one. However, analyzing results regarding the *defect lifetime* we also see no significant difference and cannot reject $H_{0t2}$. Data show that the mean time to address the first issue is about 14% faster in the treatment group if compared with the control. This is because subjects have to deal with less code that concerns a single concept, rather than having to extrapolate context information from a tangled change. At the same time the treatment group is taking longer (median) to address the second defect. We believe that this is due to the presence of two pull requests, and as such, the context switch has an overhead effect on that. From the screencast recordings we found no reviewer using multi-screen setup, therefore subjects had to close a pull-request and then review the next, where they need to gain knowledge on different code changes.

> **Result 2:** *Our experiment was not able to provide evidence for a difference in net review time between untangled pull requests (treatment) and the tangled one (control); this despite the additional overhead of dealing with two separate pull requests in the treatment group.*

## RQ3. Understanding the change's rationale

For our third research question, we seek to measure whether subjects are affected by the dependent variable in their understanding of the rationale of the change. Rationale understanding questions are Q1–Q4 (Table 4) and Fig. 2 reports the results. Q1–Q12 mark the respective questions, while answers from the (C)ontrol or (T)reatment group are marked respectively with their first letter next to the question number. Numbers in Figure count the participants' answers to questions per Likert Item. Higher scores for Q1, Q2, and Q4 mean better understanding, whereas for Q3 a lower score signifies a correct

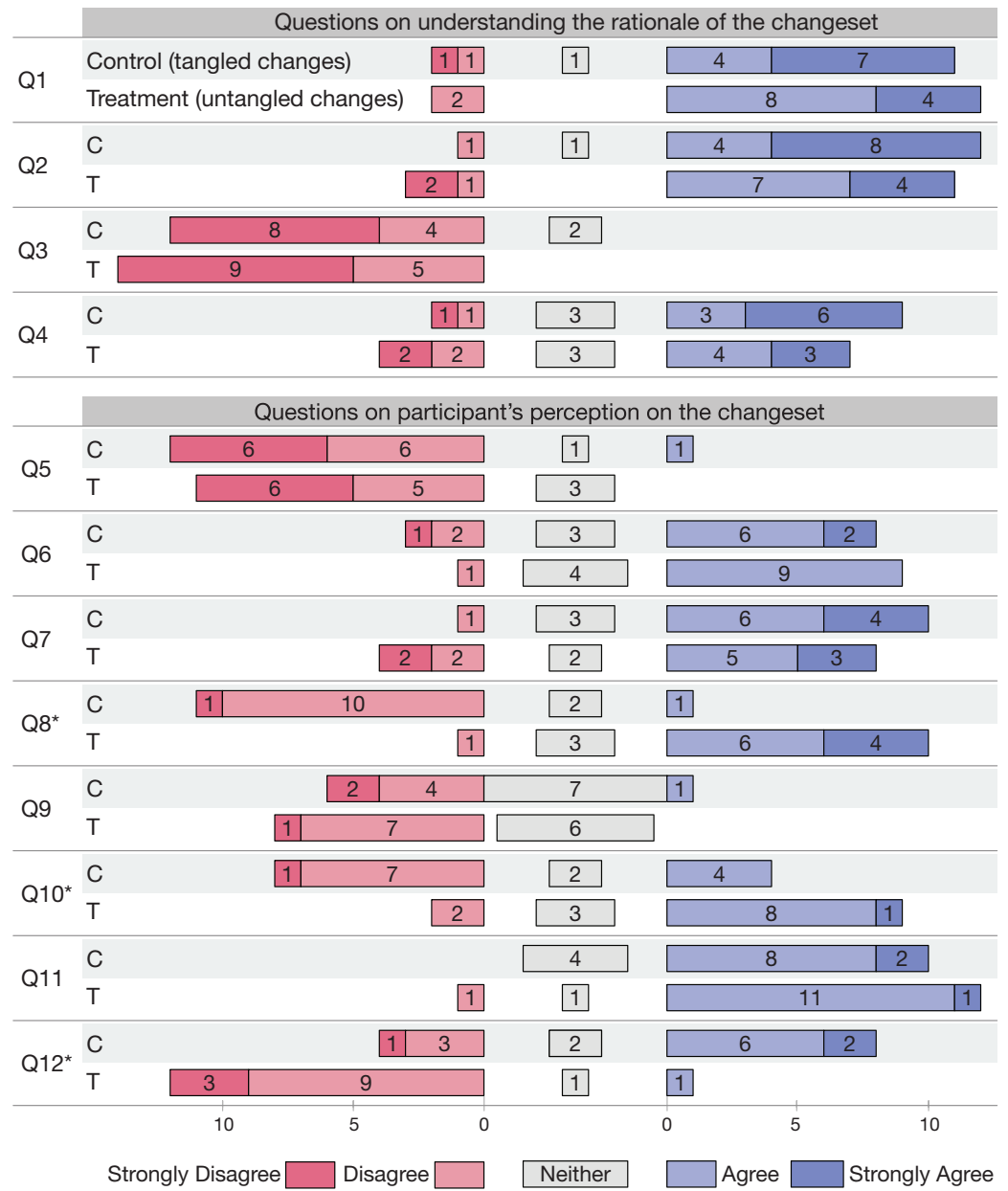

| | | Questions on understanding the rationale of the changeset | | | | | |
|---|---|---|---|---|---|---|---|
| Q1 | Control (tangled changes) | 1 1 | 1 | 4 | 7 | | |
| | Treatment (untangled changes) | 2 | | 8 | 4 | | |
| Q2 | C | 1 | 1 | 4 | 8 | | |
| | T | 2 1 | | 7 | 4 | | |
| Q3 | C | 8 4 | 2 | | | | |
| | T | 9 5 | | | | | |
| Q4 | C | 1 1 | 3 | 3 | 6 | | |
| | T | 2 2 | 3 | 4 | 3 | | |

| | | Questions on participant's perception on the changeset | | | | | |
|---|---|---|---|---|---|---|---|
| Q5 | C | 6 6 | 1 | 1 | | | |
| | T | 6 5 | 3 | | | | |
| Q6 | C | 1 2 | 3 | 6 | 2 | | |
| | T | 1 | 4 | 9 | | | |
| Q7 | C | 1 | 3 | 6 | 4 | | |
| | T | 2 2 | 2 | 5 | 3 | | |
| Q8* | C | 1 10 | 2 | 1 | | | |
| | T | 1 | 3 | 6 | 4 | | |
| Q9 | C | 2 4 | 7 | 1 | | | |
| | T | 1 7 | 6 | | | | |
| Q10* | C | 1 7 | 2 | 4 | | | |
| | T | 2 | 3 | 8 | 1 | | |
| Q11 | C | | 4 | 8 | 2 | | |
| | T | 1 | 1 | 11 | 1 | | |
| Q12* | C | 1 3 | 2 | 6 | 2 | | |
| | T | 3 9 | 1 | 1 | | | |

10  5  0        0  5  10

Strongly Disagree ▮ Disagree ▮ Neither ▯ Agree ▮ Strongly Agree ▮

**Figure 2 RQ3—Answers to questions Q1–Q12 in Table 4.** (C)ontrol and (T)reatment answers are marked with their respective first letter. Numbers count the participants' answers to questions per Likert item. Questions with * have $p < 0.05$. Full-size 🔍 DOI: 10.7717/peerj-cs.193/fig-2

understanding. As for the previous research questions, we test our hypothesis with a non-parametrical statistical test. Given the result we cannot reject the null hypothesis $H_{0u}$ of tangled pull requests reducing change understanding. Participants are in fact able to answer the questions correctly, independent of their experimental group.

After the review, our experimentation also provided a final survey (Q5–Q12 in Table 4) that participants filled in at the end. Results shown in Fig. 2 indicate that subjects judge equally the changeset (Q5), found no difficulty in understanding the changeset (Q6), agree on having understood the rationale behind the changeset (Q7). This results shows that our experiment cannot provide evidence of differences in change understanding between the two groups.

**Table 5** RQ4—Concepts from literature and their mapped keyword.

| Concept | Mapped keyword |
|---|---|
| What is the rationale behind this code change? (*Tao et al., 2012*) | Rationale |
| Is this change correct? Does it work as expected? (*Tao et al., 2012*) | Correctness |
| Who references the changed classes/methods/fields? (*Tao et al., 2012*) | Context |
| How does the caller method adapt to the change of its callees? (*Tao et al., 2012*) | Caller/callee |
| Is there a precedent or exemplar for this? (*Sillito, Murphy & De Volder, 2006*) | Similar/precedent |

Participants did not find the changeset hard to navigate (Q9), and believe that the changeset was comprehensible (Q11). Answers to questions Q9 and Q11 are surprising to us, as we would expect dissimilar results for code navigation and comprehension. In fact, change decomposition should allow subjects to navigate code easier, as well as improve source comprehension.

On the other hand, subjects from the control and treatment group judge differently when asked if the changeset was partitioned according to a logical separation of concerns (Q8), if the relationships among the changes were well structured (Q10) and if the changes were spanning too many features (Q12). These answers are in line with what we would expect, given the different structure of the code to be reviewed. The answers are different with a statistical significance for Q8, Q10 and Q12.

> **Result 3:** *Our experiment was not able to provide evidence of a difference in understanding the rationale of the changeset between the experimental groups. Subjects reviewing the untangled pull requests (treatment) recognize the benefits of untangled pull requests, as they evaluate the changeset as being (1) better divided according to a logical separation of concerns (Q8), (2) better structured (Q10), and (3) not spanning too many features (Q12).*

## RQ4. Tangled vs. untangled review patterns

For our last research question, we seek to identify differences in patterns and features during review, and their association to quantitative results. We derived such patterns from *Tao et al. (2012)* and *Sillito, Murphy & De Volder (2006)*. These two studies are relevant as they investigated the role of understanding code during the software development process. *Tao et al. (2012)* laid out a series of information needs derived from state-of-the-art research in software engineering, while *Sillito, Murphy & De Volder (2006)* focused on questions asked by professional experienced developers while working on implementing a change. The mapping found in the screencasts is shown in Table 5.

Table 6 contains the qualitative characterization, ordered by the sum of defects found. Values in each row correspond to how many times a participant in either group used that pattern to address a defect or point to a false positive.

**Table 6** RQ4—Patterns in review to address a defect or leading to a false positive.

| ID | Pattern | | | Control | | Treatment | |
|---|---|---|---|---|---|---|---|
| | 1st concept | 2nd concept | 3rd concept | Defect | FP | Defect | FP |
| P1 | Rationale | Correctness | | 8 | 3 | 4 | 0 |
| P2 | Rationale | Context | Correctness | 4 | 0 | 5 | 0 |
| P3 | Context | Rationale | Correctness | 3 | 2 | 3 | 0 |
| P4 | Context | Correctness | Caller/callee | 1 | 0 | 2 | 0 |
| P5 | Context | Correctness | | 2 | 1 | 0 | 0 |
| P6 | Correctness | Context | | 0 | 0 | 2 | 0 |
| P7 | Rationale | Correctness | Context | 0 | 0 | 1 | 0 |
| P8 | Correctness | Context | Caller/callee | 1 | 0 | 0 | 0 |
| P9 | Correctness | Context | Similar/ precedent | 1 | 0 | 0 | 1 |

Results indicate that pattern P1 is the one that led to most issues being addressed in the control group (eight), but at the same time is the most imprecise one (three false positives). We conjecture that this is related to the lack of context-seeking concept. Patterns P1 and P3 have most false positives addressed in the control group. In the treatment group, pattern P2 led to more issues being addressed (five), followed by the previously mentioned P1 (four).

Analyzing the transcribed screencasts, we note an overall trend of reviewing code changes in the control group, exploring the changeset using less context exploration than in the treatment. Among the participants belonging to the treatment, we witnessed a much more structured way of conducting the review. The overall behavior is that of getting the context of the single change, looking for the files involved, called, or referenced by the changeset, in order to grasp the rationale. All of the subjects except three repeated this step multiple times to explore a chain of method calls, or to seek for more context in that same file opening it in GitHub. We consider this the main reason to explain that untangled pull requests lead to more precise (fewer false positives) results.

> **Result 4:** *Our experiment revealed that review patterns for untangled pull requests (treatment) show more context-seeking steps, in which the participants open more referenced/related classes to review the changeset.*

# DISCUSSION

In this section, we analyze and discuss results presented in the section "Results," with consequent implications for researchers and practitioners.

## Implications for researchers

In past studies, researchers found that developers call for tool and research support for decomposing a composite change (*Tao et al., 2012*). For this reason, we were surprised that our experiment was not able to highlight differences in terms of reviewers' effectiveness (number of defects found) and reviewers' understanding of the change rationale, when the subjects were presented with smaller, self-contained changes. Further research with additional participants is needed to corroborate our findings.

If we exclude latent problems with the experiment design that we did not account for, this result may indicate that reviewers are still able to conduct their work properly, even when presented with tangled changes. However, the results may change in different contexts. For example, the cognitive load for reviewers may be higher with tangled changes, with recent research showing promising insights regarding this hypothesis (*Baum, Schneider & Bacchelli, 2019*). Therefore, the negative effects in terms of effectiveness could be visible when a reviewer has to assess a large number of changes every day, as it happens with integrators of popular projects in GitHub (*Gousios et al., 2015*). Moreover, the changes we considered are of average size and difficulty, yet results may be impacted by larger changes and/or more complex tasks. Finally, participants were not core developers of the considered software system; it is possible that core developers would be more surprised by tangled changes, find them more convoluted or less "natural," thus rejecting them (*Hellendoorn, Devanbu & Bacchelli, 2015*). We did not investigate these scenarios further, but studies can be designed and carried out to determine whether and how these aspects influence the results of the code review effort.

Given the remarks and comments of professional developers on tangled changes (*Tao et al., 2012*), we were also surprised that the experiment did not highlight any differences in the net review time between the treatment groups. Barring experimental design issues, this result can be explained by the additional context switch, which does not happen in the tangled pull request (control) because the changes are done in the same files. An alternative explanation could be that the reviewers with the untangled pull requests (treatment) spent more time "wondering around" and pinpointing small issues because they found the important defects quicker; this would be in line with the cognitive bias known as Parkinson's Law (*Parkinson & Osborn, 1957*) (all the available time is consumed). However, time to find the first and second defects (3) is the same for both experimental groups thus voiding this hypothesis. Moreover, similarly to us, *Tao & Kim (2015)* also did not find a difference with respect to time to completion in their preliminary user study. Further studies should be designed to replicate our experiment and, if results are confirmed, to derive a theory on why there is no reduction in review time.

Our initial hypothesis on why time does not decrease with untangled code changes is that reviewers of untangled changes (control) may be more willing to build a more appropriate context for the change. This behavior seems to be backed up by our qualitative analysis (section "Results"), through the context-seeking actions that we witnessed for the treatment group. If our hypothesis is not refused by further research, this could indicate that untangled changes may lead to a more thorough low-level understanding of the codebase. Despite we did not measure this in the current study, it may explain the lower number of false positives with untangled changes. Finally, untangled changes may lead to better transfer of code knowledge, one of the positive effects of code review (*Bacchelli & Bird, 2013*).

### Recommendation for practitioners

Our experiment is not able to show no negative effects when changes are presented as separate, untangled changesets, despite the fact that reviewers have to deal with two pull requests instead of one, with the subsequent added overhead and a more prominent context switch. With untangled changesets, our experiment highlighted an increased number of suggested improvements, more context-seeking actions (which, it is reasonable to assume, increase the knowledge transfer created by the review), and a lower number of wrongly reported issues.

For the aforementioned reasons, we support the recommendation that change authors prepare self-contained, untangled changeset when they need a review. In fact, untangled changesets are not detrimental to code review (despite the overhead of having more pull-requests to review), but we found evidence of positive effects. We expect the untangling of code changes to be minimal in terms of cognitive effort and time for the author. This practice, in fact, is supported by answers Q8, Q10, Q12 to the questionnaire and by comments written by reviewers in the control group (i.e., *"Please make different commit for these two features," "I would prefer having two pull requests instead of one if you are fixing two issues"*).

## CONCLUSION

The goal of the study presented in this paper is to investigate the effects of change decomposition on MCR (*Cohen, 2010*), particularly in the context of the pull-based development model (*Gousios, Pinzger & Van Deursen, 2014*).

We involved 28 subjects, who performed a review of pull request(s) pertaining to (1) a refactoring and (2) the addition of a new feature in a Java system. The control group received a single pull request with both changes tangled together, while the treatment group received two pull requests (one per type of change). We compared control and treatment groups in terms of effectiveness (number of defects found), number of false positives (wrongly reported issues), number of suggested improvements, time to complete the review(s), and level of understanding the rationale of the change. Our investigation also involved a qualitative analysis of the review performed by subjects involved in our study.

Our results suggests that untangled changes (treatment group) lead to:

1. Fewer reported false positives defects,

2. more suggested improvements for the changeset,

3. same time to review (despite the overhead of two different pull requests),

4. same level of understanding the rationale behind the change,

5. and more context-seeking patterns during review.

Our results support the case that committing changes belonging to different concepts separately should be an adopted best practice in contemporary software engineering. In fact, untangled changes not only reduce the noise for subsequent data analyses (*Herzig, Just & Zeller, 2016*), but also support the tasks of the developers in charge of reviewing the changes by increasing context-seeking patterns.

# ACKNOWLEDGEMENTS

The authors would like to thank all participants of the experiment and the pilot. We furthermore thank the fellow researchers who gave critical suggestion to help strengthening the methodology of our study.

## ADDITIONAL INFORMATION AND DECLARATIONS

### Funding

This project has received funding from the European Union's Horizon 2020 research and innovation programme under the Marie Sklodowska-Curie grant agreement No. 642954. Alberto Bacchelli has received support from the Swiss National Science Foundation through the SNF Project No. PP00P2_170529. The funders had no role in study design, data collection and analysis, decision to publish, or preparation of the manuscript.

### Grant Disclosures

The following grant information was disclosed by the authors:
European Union's Horizon 2020 research and innovation programme under the Marie Sklodowska-Curie grant agreement: 642954.
Swiss National Science Foundation through the SNF Project: PP00P2_170529.

### Competing Interests

Arie van Deursen is an Academic Editor for PeerJ Computer Science. Marco di Biase and Magiel Bruntink are employed by Software Improvement Group.

### Author Contributions

- Marco di Biase conceived and designed the experiments, performed the experiments, analyzed the data, prepared figures and/or tables, performed the computation work, authored or reviewed drafts of the paper, approved the final draft.
- Magiel Bruntink conceived and designed the experiments, authored or reviewed drafts of the paper, approved the final draft.
- Arie van Deursen conceived and designed the experiments, authored or reviewed drafts of the paper, approved the final draft.
- Alberto Bacchelli conceived and designed the experiments, authored or reviewed drafts of the paper, approved the final draft.

### Ethics

The following information was supplied relating to ethical approvals (i.e., approving body and any reference numbers):

The Human Subjects Committee of the Faculty of Economics, Business Administration and Information Technology at the University of Zurich authorized the research described in Alberto Bacchelli's research proposal with IRB 2018-024.

## Data Availability

The following information was supplied regarding data availability:

The raw data is available at: https://data.4tu.nl/repository/uuid:826f7051-35f6-4696-b648-8e56d3ea5931

# REFERENCES

**Bacchelli A, Bird C. 2013.** Expectations, outcomes, and challenges of modern code review. In: *Proceedings of the 2013 International Conference on Software Engineering, ICSE '13.* Piscataway: IEEE Press, 712–721.

**Barnett M, Bird C, Brunet J, Lahiri S. 2015.** Helping developers help themselves: automatic decomposition of code review changesets. In: *Proceedings of the 37th International Conference on Software Engineering—Volume 1, ICSE '15.* Piscataway: IEEE Press, 134–144.

**Baum T, Schneider K, Bacchelli A. 2017.** On the optimal order of reading source code changes for review. In: *2017 IEEE International Conference on Software Maintenance and Evolution (ICSME).* Piscataway: IEEE, 329–340.

**Baum T, Schneider K, Bacchelli A. 2019.** Associating working memory capacity and code change ordering with code review performance. In: *Empirical Software Engineering.* New York: Springer DOI 10.1007/s10664-018-9676-8.

**Beller M, Gousios G, Panichella A, Zaidman A. 2015.** When, how, and why developers (do not) test in their ides. In: *Proceedings of the 2015 10th Joint Meeting on Foundations of Software Engineering, ESEC/FSE 2015,* New York: ACM, 179–190.

**Benjamini Y, Hochberg Y. 1995.** Controlling the false discovery rate: a practical and powerful approach to multiple testing. *Journal of the Royal Statistical Society. Series B (Methodological)* **57(1)**:289–300 DOI 10.1111/j.2517-6161.1995.tb02031.x.

**Binkley D, Davis M, Lawrie D, Maletic J, Morrell C, Sharif B. 2013.** The impact of identifier style on effort and comprehension. *Empirical Software Engineering* **18(2)**:219–276 DOI 10.1007/s10664-012-9201-4.

**Bosu A, Greiler M, Bird C. 2015.** Characteristics of useful code reviews: an empirical study at microsoft. In: *2015 IEEE/ACM 12th Working Conference on Mining Software Repositories.* Piscataway: IEEE, 146–156.

**Cliff N. 1993.** Dominance statistics: ordinal analyses to answer ordinal questions. *Psychological Bulletin* **114(3)**:494–509 DOI 10.1037/0033-2909.114.3.494.

**Cohen J. 1992.** Statistical power analysis. *Current Directions in Psychological Science* **1(3)**:98–101 DOI 10.1111/1467-8721.ep10768783.

**Cohen J. 2010.** Modern code review. In: Oram A, Wilson G, eds. *Making Software.* Chapter 18. Sebastopol: O'Reilly, 329–338.

**Cook TD, Campbell DT. 1979.** *Quasi-experimentation: design and analysis for field settings.* Vol. 3. Chicago: Rand McNally.

**Cowan N. 1998.** *Attention and memory: an integrated framework.* Vol. 26. Oxford: Oxford University Press.

**di Biase M, Bruntink M, Bacchelli A. 2016.** A security perspective on code review: the case of chromium. In: *Proceedings of the 16th IEEE International Working Conference on Source Code Analysis and Manipulation, SCAM 2016, October 2-3, 2016.* Piscataway: IEEE, 21–30.

**di Biase M, Bruntink M, Van Deursen A, Bacchelli A. 2018.** The effects of change decomposition on code review—a controlled experiment—online appendix. *Available at https://data.4tu.nl/repository/uuid:826f7051-35f6-4696-b648-8e56d3ea5931.*

**Dias M, Bacchelli A, Gousios G, Cassou D, Ducasse S. 2015.** Untangling fine-grained code changes. In: *Proceedings of the 22nd International Conference on Software Analysis, Evolution, and Reengineering, SANER 2015.* Piscataway: IEEE Computer Society, 341–350.

**Gousios G, Pinzger M, Van Deursen A. 2014.** An exploratory study of the pull-based software development model. In: *Proceedings of the 36th International Conference on Software Engineering—ICSE 2014, (May 2014).* New York: ACM, 345–355.

**Gousios G, Zaidman A, Storey M, Van Deursen A. 2015.** Work practices and challenges in pull-based development: the integrator's perspective. In: *Proceedings of the 37th International Conference on Software Engineering—Volume 1, ICSE '15.* Piscataway: IEEE Press, 358–368.

**Hellendoorn VJ, Devanbu PT, Bacchelli A. 2015.** Will they like this? Evaluating code contributions with language models. In: *Proceedings of the 12th Working Conference on Mining Software Repositories.* Piscataway: IEEE Press, 157–167.

**Herzig K, Just S, Zeller A. 2016.** The impact of tangled code changes on defect prediction models. *Empirical Software Engineering* **21(2)**:303–336 DOI 10.1007/s10664-015-9376-6.

**Herzig K, Zeller A. 2013.** The impact of tangled code changes. In: *Mining Software Repositories (MSR) '13 Proceedings of the 10th IEEE Working Conference on Mining Software.* Piscataway: IEEE, 121–130.

**James W. 2013.** *The principles of psychology.* Redditch: Read Books Ltd.

**Kirinuki H, Higo Y, Hotta K, Kusumoto S. 2014.** Hey! are you committing tangled changes? In: *Proceedings of the 22nd International Conference on Program Comprehension, ICPC 2014.* New York: ACM, 262–265.

**Ko A, LaToza T, Burnett M. 2015.** A practical guide to controlled experiments of software engineering tools with human participants. *Empirical Software Engineering* **20(1)**:110–141 DOI 10.1007/s10664-013-9279-3.

**Mayring P. 2000.** Qualitative content analysis. *Forum: Qualitative Social Research* **1(2)**:159–176 DOI DOI 10.17169/fqs-1.2.1089.

**McIntosh S, Kamei Y, Adams B, Hassan A. 2014.** The impact of code review coverage and code review participation on software quality: a case study of the qt, vtk, and itk projects. In: *Proceedings of the 11th Working Conference on Mining Software Repositories, MSR 2014.* New York: ACM, 192–201.

**McIntosh S, Kamei Y, Adams B, Hassan AE. 2016.** An empirical study of the impact of modern code review practices on software quality. *Empirical Software Engineering* **21(5)**:2146–2189 DOI 10.1007/s10664-015-9381-9.

**Morales R, McIntosh S, Khomh F. 2015.** Do code review practices impact design quality? A case study of the Qt, Vtk, and Itk projects. In: *Proceedings of the 22nd International Conference on Software Analysis, Evolution and Reengineering, SANER 2015.* Piscataway: IEEE, 171–180.

**Murphy-Hill E, Parnin C, Black A. 2012.** How we refactor, and how we know it. *IEEE Transactions on Software Engineering* **38(1)**:5–18 DOI 10.1109/tse.2011.41.

**Oppenheim A. 2000.** *Questionnaire design, interviewing and attitude measurement.* London: Bloomsbury Publishing.

**Parkinson CN, Osborn RC. 1957.** *Parkinson's law, and other studies in administration.* Vol. 24. Boston: Houghton Mifflin.

**Perneger TV. 1998.** What's wrong with bonferroni adjustments. *British Medical Journal* **316(7139)**:1236–1238 DOI 10.1136/bmj.316.7139.1236.

**Prechelt L, Tichy W. 1998.** A controlled experiment to assess the benefits of procedure argument type checking. *IEEE Transactions on Software Engineering* **24(4)**:302–312 DOI 10.1109/32.677186.

**Ram A, Sawant AA, Castelluccio M, Bacchelli A. 2018.** What makes a code change easier to review: an empirical investigation on code change reviewability. In: *Proceedings of the 2018 26th ACM Joint Meeting on European Software Engineering Conference and Symposium on the Foundations of Software Engineering (ESEC/FSE 2018)*. New York, NY: ACM, 201–202 DOI 10.1145/3236024.3236080.

**Rigby PC, Bird C. 2013.** Convergent contemporary software peer review practices. In: *Proceedings of the 2013 9th Joint Meeting on Foundations of Software Engineering, ESEC/FSE 2013*. New York: ACM, 202–212.

**Rigby P, Cleary B, Painchaud F, Storey M, German D. 2012.** Contemporary peer review in action: lessons from open source development. *IEEE Software* 29(6):56–61 DOI 10.1109/ms.2012.24.

**Rigby P, German D, Cowen L, Storey M. 2014.** Peer review on open-source software projects. *ACM Transactions on Software Engineering and Methodology* 23(4):1–33.

**Romano J, Kromrey J, Coraggio J, Skowronek J. 2006.** Appropriate statistics for ordinal level data: should we really be using t-test and cohen'sd for evaluating group differences on the nsse and other surveys. In: *Annual Meeting of the Florida Association of Institutional Research*, 1–33.

**Sadowski C, Söderberg E, Church L, Sipko M, Bacchelli A. 2018.** Modern code review: a case study at google. In: *Proceedings of the 40th International Conference on Software Engineering Software Engineering: in Practice (ICSE-SEIP '18)*. New York, NY: ACM, 181–190 DOI 10.1145/3183519.3183525.

**Schreier M. 2013.** Qualitative content analysis. In: Flick U, ed. *The SAGE Handbook of Qualitative Data Analysis*. London: SAGE, 170–183.

**Sharif B, Falcone M, Maletic JI. 2012.** An eye-tracking study on the role of scan time in finding source code defects. In: *Proceedings of the Symposium on Eye Tracking Research and Applications*. New York: ACM, 381–384.

**Shiffrin RM. 1988.** Attention. In: Atkinson RC, Herrnstein RJ, Lindzey G, Luce RD, eds. *Stevens' Handbook of Experimental Psychology: Perception and Motivation; Learning and Cognition*. Vol. 2. Oxford: John Wiley & Sons, 739–811.

**Sillito J, Murphy G, De Volder K. 2006.** Questions programmers ask during software evolution tasks. In: *Proceedings of the 14th ACM SIGSOFT International Symposium on Foundations of Software Engineering*. New York: ACM, 23–34.

**Slavin R. 1987.** Mastery learning reconsidered. *Review of Educational Research* 57(2):175–213 DOI 10.3102/00346543057002175.

**Tao Y, Dang Y, Xie T, Zhang D, Kim S. 2012.** How do software engineers understand code changes? An exploratory study in industry. In: *Proceedings of the ACM SIGSOFT 20th International Symposium on the Foundations of Software Engineering, FSE '12*, New York: ACM, 1–11.

**Tao Y, Kim S. 2015.** Partitioning composite code changes to facilitate code review. In: *Proceedings of the 12th Working Conference on Mining Software Repositories*. Piscataway: IEEE, 180–190.

**Thongtanunam P, McIntosh S, Hassan AE, Iida H. 2017.** Review participation in modern code review. *Empirical Software Engineering* 22(2):768–817 DOI 10.1007/s10664-016-9452-6.

**Uwano H, Nakamura M, Monden A, Matsumoto K. 2006.** Analyzing individual performance of source code review using reviewers' eye movement. In: *Proceedings of the 2006 Symposium on Eye Tracking Research & Applications*. New York: ACM, 133–140.

**Wickens CD. 1991.** Processing resources and attention. *Multiple-Task Performance*. London: Taylor & Francis, 3–34.

**Wohlin C, Runeson P, Höst M, Ohlsson M, Regnell B, Wesslén A. 2012.** *Experimentation in software engineering*. Berlin/Heidelberg: Springer Science & Business Media.